\documentclass{article}

\usepackage{arxiv}

\usepackage[utf8]{inputenc} 
\usepackage[T1]{fontenc}    
\usepackage{xurl}            
\usepackage{booktabs}       
\usepackage{amsfonts}       
\usepackage{nicefrac}       
\usepackage{microtype}      
\usepackage{lipsum}		
\usepackage{graphicx}
\usepackage{natbib}
\usepackage{doi}
\usepackage{amsmath,amssymb}%
\usepackage{placeins}
\usepackage{hyperref}       

\title{SCALE: Scientific Concept Aggregation via LLMs and Embeddings for Fine-Grained Taxonomy Extension}

\author{
{\normalfont Daniele Raimondi}\\
\texttt{daniele.raimondi@mdpi.com}
\and
{\normalfont Feichi Lu}\\
\texttt{feichi.lu@mdpi.com}
\and
{\normalfont Oliver Grun}\\
\texttt{oliver.grun@mdpi.com}
\and
{\normalfont Mariia Eremina}\\
\texttt{mariia.eremina@mdpi.com}
\and
{\normalfont Andrea Perlato\thanks{Corresponding author.}}\\
\texttt{andrea.perlato@mdpi.com}
}

\date{MDPI Data Intelligence Team \\
MDPI AG \\
Grosspeteranlage 5 \\
CH-4052 Basel, Switzerland 
 }

\renewcommand{\headeright}{}

\hypersetup{
pdftitle={A template for the arxiv style},
pdfsubject={q-bio.NC, q-bio.QM},
pdfauthor={David S.~Hippocampus, Elias D.~Striatum},
pdfkeywords={First keyword, Second keyword, More},
}

\begin{document}
\maketitle

\begin{abstract}
	The increasing specialization of scientific research challenges existing
classification systems, which provide effective representations of broad
disciplines and research topics but often fail to capture the fine-grained
conceptual structure of contemporary science. Author keywords offer greater
specificity, but their fragmentation, redundancy, and terminological variability
limit their use as stable units of knowledge organization. We introduce SCALE
(Scientific Concept Aggregation via LLMs and Embeddings), a framework that
extends the OpenAlex taxonomy with a new level of scientific Concepts below
Topics. Rather than treating keywords as isolated descriptors, SCALE organizes
semantically related terms into coherent and interpretable conceptual units and
integrates them within the existing disciplinary hierarchy. The framework
combines scientific text embeddings, large language models, and graph-based
community detection to construct this additional layer at scale. The resulting
taxonomy enables scientific literature to be read through an intermediate
conceptual level between broad research topics and individual documents. This
perspective provides a more detailed representation of how scientific knowledge
is structured, specialized, and connected across disciplines. By transforming
heterogeneous author terminology into reusable hierarchical units, SCALE offers
a foundation for fine-grained scholarly classification, scientometric analysis,
research monitoring, and future ontology development.
\end{abstract}

\keywords{Scientific taxonomy\and Large language models\and Text embeddings\and Graph clustering\and Knowledge organization}

\section{Introduction}
\label{sec:introduction}

The rapid expansion of scholarly literature has made organizing research increasingly difficult. Scientific output continues to grow in volume, while research fields become more specialized and new interdisciplinary areas emerge at the boundaries of established disciplines \citep{bornmann2015}. As a result, simply assigning publications to broad disciplinary categories is no longer sufficient. Organizing research has become essential for information retrieval, scientometric analysis, the detection of emerging trends, and evidence-informed decision-making.

Research communities have proposed several approaches to make the structure of science more visible and easier to analyze. Classification systems, citation-based methods, term co-occurrence analysis, knowledge graphs, ontologies, and semantic representations have each contributed to this goal from different perspectives \citep{fortunato2018}. Large-scale infrastructures such as OpenAlex show how hierarchical structures can connect disciplines, subdisciplines, and research topics across the global scholarly record \citep{priem2022}. These systems provide an important foundation for navigating and analyzing scientific literature at scale.

However, a key limitation remains. Existing hierarchical classifications represent broad domains and research topics well, yet they rarely offer enough granularity to describe highly specialized areas of contemporary research \citep{clarivate_wos_categories, elsevier_scopus_asjc, abs_anzsrc_2020}. Author keywords offer a more detailed view of individual publications, but they are fragmented, redundant, and strongly affected by terminological variation. The same concept may appear under different names, while similar terms may carry different meanings across fields. This makes author keywords unreliable as stable units at scale.

These limitations create a gap between high-level classifications and the highly specific descriptions associated with individual documents. What is needed is a new level, positioned below existing topics, specific enough to capture specialized research themes but still stable enough to fit into a coherent hierarchy. Such a level should also be scalable and compatible with existing classification infrastructures, rather than requiring a completely new system.

In this work, we introduce Concepts as a new hierarchical level designed to bridge this gap. They form a new level below existing OpenAlex Topics: coherent groups of related terms, built from the raw vocabulary of author keywords but organized into a far more stable and reusable structure. To construct them, we propose a scalable framework that combines semantic representations, graph-based clustering, and automated procedures for annotating and integrating them into the hierarchy. Applied to the OpenAlex taxonomy, the framework extends the existing structure with approximately 114,000 Scientific Concepts connected to the Topics already present in the OpenAlex hierarchy \citep{openalexTopics}.

The resulting taxonomy provides a basis for future applications in scientometric analysis, semantic exploration, and large-scale research intelligence. The next section reviews the relevant literature and clarifies the methodological context that led to the development of Scientific Concepts.


\section{Related Work}
\label{sec:relatedwork}
This section reviews prior work relevant to the construction of Concepts, organized around four lines of research: science mapping, semantic representation of scientific text, topic modeling, and knowledge organization systems. It then examines OpenAlex specifically, before identifying the gap that Concepts are designed to address.

\subsection{Mapping and Visualization of Science}
Understanding the structure of science has long been a central concern in scientometrics. Science mapping methods aim to identify, visualize, and analyze relationships between disciplines, research areas, and topics, transforming large bibliographic collections into interpretable representations of research activity \citep{petrovich2021}.

A major line of work relies on citation-based methods, bibliographic coupling, and term co-occurrence analysis to reveal emerging structures and thematic relationships within the literature. Tools such as VOSviewer have contributed significantly to the adoption of these approaches by enabling the construction of science maps based on bibliometric and semantic similarities \citep{vanEck2010}. More recent comparative approaches have extended this perspective by supporting systematic comparisons of different representations of scientific structure and their evolution over time.

These methods are valuable for describing and visualizing how research areas are connected. Still, they are usually designed to reveal patterns, clusters, or relationships in bibliographic data rather than to create persistent conceptual units that a hierarchical classification system can incorporate. Science mapping therefore provides an important foundation for understanding the structure of research, without addressing the need for stable, fine-grained, hierarchically organized units.

\subsection{Semantic Representation of Scientific Literature}
Advances in semantic representation have changed the way scientific literature is analyzed and organized. Traditional approaches relied mainly on keywords, term frequencies, and co-occurrence patterns. More recent developments in Natural Language Processing have introduced models that represent scientific documents as dense vectors in high-dimensional semantic spaces, enabling the capture of relationships among publications beyond lexical similarity.

Transformer-based models such as SPECTER have played an important role in this
development. SPECTER introduced document representations that combine textual
information with citation-based signals. By embedding related articles close to
one another in a shared vector space, these representations have supported tasks
such as information retrieval, document classification, and large-scale analysis
of scientific literature \citep{cohan2020}.

Semantic representation models provide a powerful technological basis for analyzing scientific literature at scale. Still, these models primarily represent documents or concepts in a semantic space rather than construct explicit structures for knowledge organization. They can capture meaningful relationships among publications without directly producing an interpretable, maintainable conceptual hierarchy.

\subsection{Topic Modeling Approaches}
Researchers have long used topic modeling to discover latent thematic structure in large text collections. Classical approaches such as Latent Dirichlet Allocation (LDA) represent documents as mixtures of topics, where a distribution over words defines each topic, without requiring predefined categories. More recent neural approaches, such as BERTopic, combine transformer-based embeddings with clustering to produce topics that better capture semantic similarity beyond word co-occurrence \citep{blei2003,grootendorst2022}.

These methods are effective at uncovering patterns in unlabeled corpora. Still, the resulting topics typically remain corpus-specific, can vary across different runs, and do not align with any pre-existing classification hierarchy. As a result, they do not directly produce reusable, hierarchically organized units for large-scale scholarly classification.

\subsection{Knowledge Organization Systems and Scientific Ontologies}
Alongside advances in semantic representation, researchers have developed explicit frameworks to organize concepts and their relationships. Knowledge Organization Systems (KOS) include taxonomies, ontologies, and classification schemes that represent knowledge in a structured and interpretable form.

Recent computational approaches have addressed the automatic construction and 
completion of topic taxonomies. TaxoGen constructs corpus-specific topic 
hierarchies by combining adaptive term embeddings with recursive clustering of 
semantically coherent terms, whereas TaxoCom uses a partial taxonomy to guide 
the discovery of both existing and novel subtopic clusters. More recently, 
TaxoAdapt combines LLM-based multidimensional taxonomy generation with 
corpus-grounded hierarchical classification and iterative expansion to represent 
evolving scientific domains 
\citep{zhang2018taxogen,lee2022taxocom,kargupta2025taxoadapt}. 
Unlike these approaches, our framework does not construct or recursively complete 
an entire corpus-specific hierarchy. Instead, it preserves the four-level OpenAlex 
taxonomy and introduces a controlled fifth level of fine-grained Concepts below 
Topics through LLM-based granularity and field classification, field-specific 
semantic graphs, and Leiden clustering.

In the scientific domain, the Computer Science Ontology (CSO) provides an important example of how large-scale representations of research areas can combine automated methods with expert knowledge. Related work, such as Klink-2, has demonstrated how computational methods can identify, link, and update relationships among scientific topics over time, supporting the construction of dynamic semantic research networks \citep{salatino2020,osborne2015}.

Even so, combining broad disciplinary coverage, fine granularity, and continuous updating within a single system remains difficult in practice.

\subsection{{OpenAlex} and the Challenge of Fine-Grained Classification}
OpenAlex is one of the largest infrastructures for organizing and classifying the global scholarly record. Its dataset is released under the CC0 license, allowing unrestricted reuse \citep{openalexPricing}. The OpenAlex taxonomy is structured into Domains, Fields, Subfields, and Topics, providing a broad and coherent representation of contemporary research areas \citep{openalexTopics}.

Despite its scale and utility, the OpenAlex hierarchy still leaves the challenge of representing more granular research themes open. Earlier concept-based approaches inherited from the Microsoft Academic Graph (MAG) showed how difficult it is to maintain conceptual units that remain stable, interpretable, and consistent over time \citep{wang2020mag}. For this reason, OpenAlex deprecated Concepts in favor of the current Topics hierarchy, complemented by an auxiliary entity called Keywords, derived from each topic \citep{openalexConcepts}.

These OpenAlex keywords, however, are a fixed, closed set (currently over 26,000, ten per topic) generated to describe existing topics and assigned to individual works for retrieval purposes; they do not constitute a hierarchical level of stable, reusable concept units with an explicit path to subfields, fields, and domains \citep{openalexKeywords}.

This choice reflects a broader limitation of large-scale scholarly classification. Existing hierarchical levels often remain too general for highly specialized research areas. In contrast, keyword- and similarity-based approaches provide greater detail but do not necessarily produce a stable and interpretable structure. As a result, existing systems do not produce a coherent, maintainable layer beneath the Topics hierarchy from this material.

\subsection{Toward Concepts}
The literature reviewed in the previous sections shows a clear progression in the methods used to represent and organize scientific literature. Science mapping has provided tools for identifying and visualizing the structure of research. Semantic representation models have improved the ability to capture conceptual relationships among documents, while Knowledge Organization Systems have introduced explicit and interpretable structures for organizing concepts.

However, each of these approaches addresses only part of the problem: detecting structure, representing meaning, or organizing concepts, but rarely all three at once. Concepts address this gap without replacing existing taxonomies, ontologies, or classification infrastructures. Rather, they extend the hierarchy downward, increasing the granularity of scientific classification while remaining coherent, interpretable, and easy to maintain.

\section{Methods}
\label{sec:methods}

This section describes our framework for constructing a fifth Concept level and integrating it below the existing OpenAlex taxonomy. The core challenge is to transform noisy, heterogeneous author keywords into stable and reusable scientific Concepts that are fine-grained but not too specific for specialized research themes.  

\begin{figure}[!htbp]
    \centering
    \includegraphics[width=0.70\textwidth,keepaspectratio]{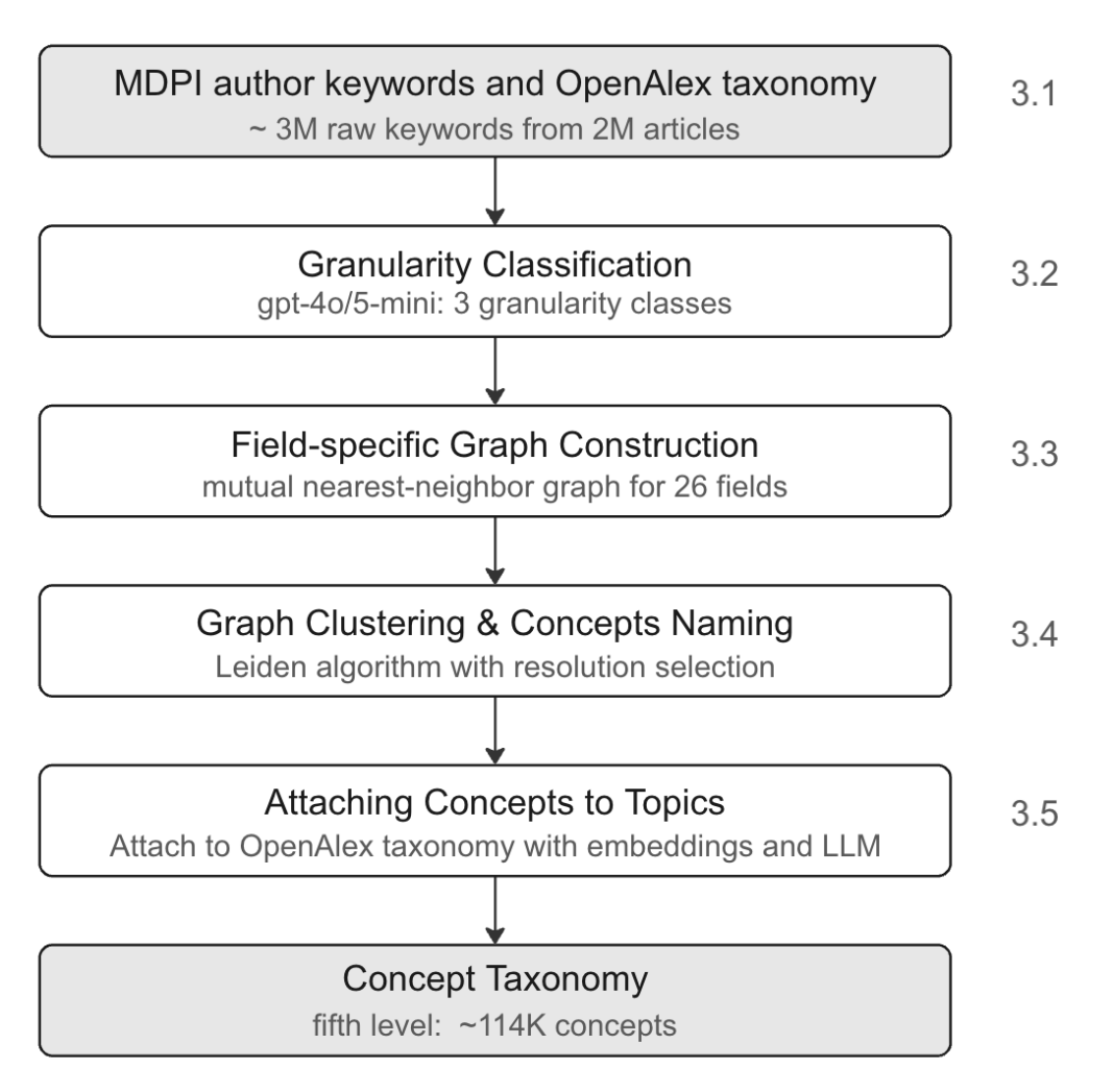}
    \caption{Overview of the Concepts framework. Author keywords are progressively transformed into Concepts through semantic processing, clustering, and hierarchical integration with the OpenAlex taxonomy.}
    \label{fig:workflow}
\end{figure}

Starting from MDPI author keywords introduced in Section \ref{subsec:data-sources}, our framework addresses the challenging problem through four main steps. In Section \ref{subsec:granularity-classification}, we identify terms that match the target level of granularity for a fifth-level taxonomy. In Section \ref{subsec:construct-graph}, we construct field-specific semantic graphs for keyword community detection. Section \ref{subsec:graph-clustering} focuses on graph clustering techniques and concept renaming. In Section \ref{subsec:attaching}, we attach resulting Concepts to OpenAlex Topics. Additionally, Section \ref{subsec:taxonomy-maintenance} defines an incremental maintenance procedure that allows new Concepts to be added over time. Fig.~\ref{fig:workflow} summarizes the complete methodological workflow. 

\subsection{Data Sources: OpenAlex Topics and MDPI Author Keywords}
\label{subsec:data-sources}

\textbf{The OpenAlex Taxonomy} organizes scientific knowledge into 4 levels: Domains, Fields, Subfields, and Topics. In this work, Topics \citep{openalexTopics} serve as the parent level for the new concepts to be attached.

\textbf{MDPI author keywords} include approximately three million distinct keywords extracted from nearly two million scientific articles from MDPI \citep{mdpi2026}. The data provide a broad, detailed, and multidisciplinary representation of specialized research concepts and methods, making it a suitable starting point for constructing a fine-grained scientific taxonomy level.

\subsection{Identifying the Target Level of Granularity}
\label{subsec:granularity-classification}

A central design decision is to determine which author keywords are suitable to be included as fifth-level Concepts. The target level should be narrower than existing OpenAlex Topics, but broader than too specific entities, such as products, biological species, genes, chemical compounds, and long contextual phrases. The goal is therefore to identify reusable scientific classification units rather than to preserve every keyword that appears in the source data.

The author keywords are then classified according to their level of granularity. Each keyword is assigned to one of three categories: high-level, Concept-level, or low-level. High-level terms include existing OpenAlex 4 levels, examples such as Computer Science, Physics, Artificial Intelligence, and Epidemiology. Low-level terms are those too specific or contextual to serve as stable taxonomy units. Concept-level terms represent the target granularity of the fifth level, including methods, techniques, research themes, and specialized application areas. We use OpenAI models to classify the keywords into different conceptual granularity \citep{gpt4o_mini, gpt5}.

Only keywords classified as Concept-level are retained for subsequent field-specific clustering. This filtering step ensures that the resulting taxonomy is not a simple keyword inventory, but a controlled layer of scientific Concepts positioned between OpenAlex Topics and raw author keywords.

\subsection{Construction of Field-Specific Semantic Graph}
\label{subsec:construct-graph}

After identifying candidate Concept-level keywords, the next step is to detect keyword communities with similar meanings that can form the basis of the Concept layer. To do so, we first construct a semantic graph over the candidate keywords, where nodes represent keywords and edges represent semantic relationships between them. This graph-based strategy is well suited to scientific terminology because conceptual communities can vary substantially in density, size, and internal structure. Unlike direct clustering approaches, which often impose stronger assumptions about cluster shape or scale, graph clustering allows Concepts to emerge from local semantic relationships among keywords \citep{ester1996density, traag2019}.

However, a global semantic graph may ignore disciplinary context. Author keywords do not always carry a single stable meaning across all fields: the same or similar keyword may refer to different concepts in different disciplines, and keywords from different Fields may appear close in embedding space even when they belong to distinct research contexts. To preserve semantic coherence, we therefore construct field-specific semantic graphs rather than a single global graph. Each candidate keyword is assigned to one or more of the 26 OpenAlex Fields using an LLM \citep{gpt4o_mini}. When a keyword is relevant to more than one discipline, it is processed independently within each assigned Field, rather than being forced into a single disciplinary path.

To build the semantic graph for each field, we first enrich the contexts of the author keywords with a short field-specific definition generated by an LLM. The keyword and its definition are then jointly encoded with SPECTER2 \citep{singh2023}, producing a contextualized semantic representation that captures both the keyword itself and its disciplinary meaning. To create semantic edges, we perform nearest-neighbor search using the efficient HNSW algorithm \citep{malkov2020}, which identifies related terms in high-dimensional embedding space. To reduce weak or accidental connections, two keyword nodes are linked only when they appear in each other's nearest-neighbor lists, and their semantic similarity exceeds a predefined threshold.
\subsection{Graph Clustering Towards Concept Layer}
\label{subsec:graph-clustering}

Based on the field-specific semantic graph, we apply the Leiden algorithm \citep{traag2019} to identify semantic communities. These communities group together closely related keywords and provide the empirical basis for constructing the final Concepts. A key challenge in this step is controlling the granularity of the clusters. If clusters are too large, the resulting Concepts may become overly general; if clusters are too small, the resulting Concepts may become too specific and fragmented. Therefore, clustering parameters, especially the Leiden resolution parameter, must be tuned to obtain an appropriate level of conceptual granularity.

To address this trade-off, we manually annotate two sample sets of Concept pairs: must-link pairs, which should be placed in the same cluster, and should-not-link pairs, which should be separated into different clusters. We then perform a grid search over the graph construction and clustering parameters, and select the configuration that achieves the best balance between must-link precision, should-not-link precision, and a reasonable total number of clusters. The selected configuration and its evaluation are reported in Section \ref{subsec:clustering-optimization-hierarchical-integration}.

The semantic communities detected by the Leiden algorithm provide the structural basis for constructing the final Concept layer. However, these communities are initially only clusters of related candidate keywords and must be converted into interpretable taxonomy units. For each cluster, we prompt an LLM with the constituent candidate keywords and their field-specific definitions to generate a representative Concept name and a short description. The prompt is designed to avoid overly broad labels and to ensure that the generated name closely reflects the shared meaning of the constituent terms. Through this step, each Leiden community is formalized as a named and described Concept, producing the fine-grained Concept layer that is subsequently attached below OpenAlex Topics.

\subsection{Attaching Concepts to OpenAlex Topics}
\label{subsec:attaching}

After name and description assignment, each Concept is attached to OpenAlex Topics. First, the name and description are used to generate a vector representation of the Concept \citep{singh2023}. This representation is then compared with the vector representations of OpenAlex Topics using cosine similarity, and only the most similar Topics are retained as candidates for hierarchical attachment. This retrieval step narrows the search space before LLM evaluation, avoiding the need to ask the LLM to consider all OpenAlex Topics. GPT-4o-mini then evaluates this limited candidate set and selects the most appropriate Topic based on the Concept and disciplinary context. When a Concept matches more than one Topic above a similarity threshold, all matching Topics are retained rather than only the single best one.

\begin{figure}[!htbp]
    \centering
    \includegraphics[width=\textwidth,keepaspectratio]{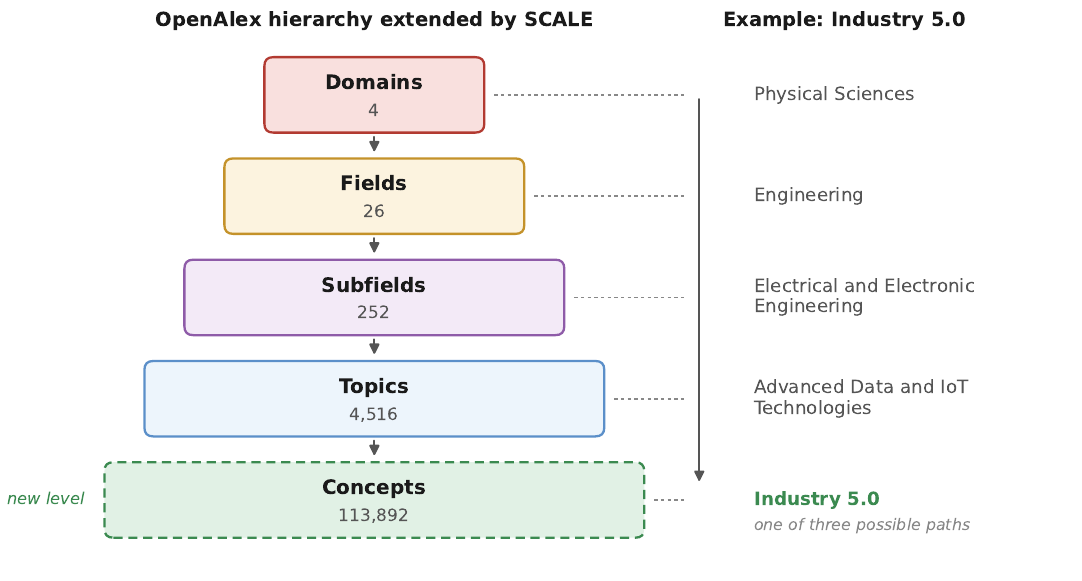}
    \caption{Five-level taxonomy extending the OpenAlex hierarchy with Concepts.}
    \label{fig:five-level-taxonomy}
\end{figure}

The selected links are finally checked to reduce redundant or incoherent associations. At the end of this stage, each Concept is connected to one or more OpenAlex Topics and receives a complete path within the taxonomy. In this way, the OpenAlex hierarchy is extended to include a fifth level comprising approximately 114,000 Concepts. Fig.~\ref{fig:five-level-taxonomy} illustrates the final five-level taxonomy and, using Industry 5.0 as an example, shows how a Concept is integrated into the OpenAlex hierarchy.

\subsection{Taxonomy Maintenance}
\label{subsec:taxonomy-maintenance}
Scientific literature evolves rapidly, so the taxonomy must be updated to reflect emerging scientific trends. Our maintenance procedure starts by identifying emerging keywords from the MDPI keywords in Section \ref{subsec:data-sources}. A keyword is defined as emerging when the gain of its recent frequency percentile relative to its historical percentile exceeds a fixed threshold. We then conduct similarity comparisons both among emerging keywords and between emerging keywords and existing Concepts, in order to filter out redundant terms and identify keywords that may serve as new Concepts in the taxonomy.

The remaining candidates are processed through a granularity filtering step similar to Section~\ref{subsec:granularity-classification}. Candidate Concepts with proper granularity are then attached to OpenAlex Topics following Section~\ref{subsec:attaching}. Each new Concept is tagged with a version label identifying the update in which it was introduced. This incremental approach keeps the taxonomy up to date while preserving the stability of the existing structure and providing a foundation for future applications such as article classification.

\section{Results}
\label{sec:results}
This section reports the selected graph-clustering configuration, its performance on curated must-link and should-not-link pairs, the resulting Concept layer, and its integration into the OpenAlex hierarchy. The Taxonomy Explorer and Atlantis interfaces are then presented, together with the public release of the taxonomy dataset. Downstream evidence of operational use and expert assessment of Concept applicability is also reported.

\subsection{Clustering Optimization and Hierarchical Integration}
\label{subsec:clustering-optimization-hierarchical-integration}
The quality of the generated Concepts depends on the parameters that control graph construction and clustering. To identify the final configuration, the optimization procedure tested three parameters through grid search: the number of nearest neighbors, the minimum similarity threshold, and the Leiden resolution parameter.

The evaluation used manually curated pairs of candidate terms. Must-link pairs
represented terms expected to belong to the same semantic community, whereas
should-not-link pairs represented terms expected to belong to different
communities. The pairs focused on difficult boundary cases observed during
clustering development, rather than on obvious synonym or unrelated pairs.
Must-link precision was defined as the proportion of must-link pairs assigned to
the same community. In contrast, should-not-link precision was defined as the
proportion of should-not-link pairs assigned to different communities. Overall accuracy was defined as the proportion of correctly classified pairs among all evaluated pairs.

\begin{figure}[!htbp]
    \centering
    \includegraphics[width=\textwidth,keepaspectratio]{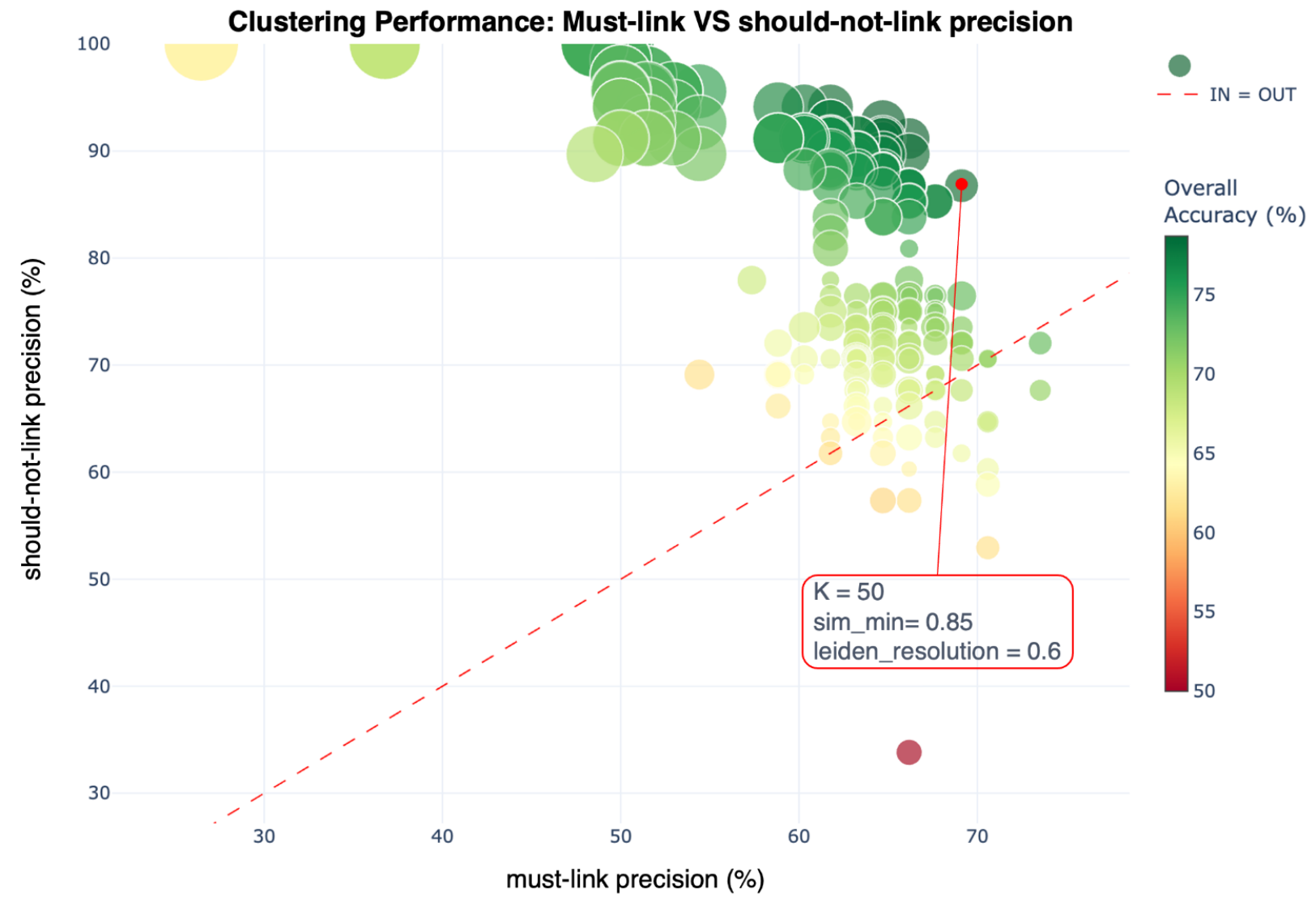}
    \caption{Performance of the clustering configurations evaluated during grid search. The highlighted point represents the final parameter configuration adopted by the framework. $K$: the number of nearest neighbors, $\text{sim\_min}$: the minimum similarity threshold, $\text{leiden\_resolution}$: the Leiden resolution parameter.}
    \label{fig:grid-search}
\end{figure}

Fig.~\ref{fig:grid-search} reports the results of the grid search. Each point
represents a specific parameter configuration, with must-link precision on the
x-axis and should-not-link precision on the y-axis; bubble size indicates the
number of detected communities, corresponding to candidate Concepts. The dashed
line marks configurations in which the two precision values are equal. The final
configuration was selected by jointly considering the two precision scores and
the resulting number of communities. The highlighted point represents the
configuration adopted for constructing the taxonomy.

\FloatBarrier

Based on this evaluation of 68 must-link and 68 should-not-link pairs curated by the authors, the final configuration used k=50, a minimum similarity threshold of 0.85, and a Leiden resolution parameter of 0.6, achieving 69.5\% must-link precision and 87.5\% should-not-link precision, serving as a consistency check on challenging aggregation and separation decisions rather than a random-sample benchmark of clustering quality. The subsequent
clustering and naming stages produced 113,892 Concepts, each representing a specific research theme formed by grouping semantically related candidate concepts within the corresponding Field.



After hierarchical attachment, at least one Concept was associated with 98.8\%
of the 4,516 OpenAlex Topics (4,461 Topics). Because each OpenAlex Topic belongs to a Subfield, Field, and Domain, this attachment gives each Concept a complete position within the OpenAlex hierarchy. The resulting structure extends OpenAlex with a fifth level while preserving its original organization. When semantically appropriate, the taxonomy can place a Concept under more than one Topic, allowing interdisciplinary research themes to appear across multiple disciplinary paths.

At the Domain level, the 113,892 Concepts have the following distribution across the
OpenAlex hierarchy: Physical Sciences accounted for 65,615 Concepts (57.6\%),
followed by Social Sciences with 18,814 (16.5\%), Health Sciences with 15,437
(13.6\%), and Life Sciences with 14,026 (12.3\%).

\subsection{Public Release and Exploration Tools}
\label{subsec:taxonomy-access-exploration}
The generated taxonomy is made available through three access channels: the
Taxonomy Explorer for hierarchical navigation, Atlantis for map-based
exploration of Concepts, and an open dataset containing the full hierarchy and
Concept descriptions.

\subsubsection{Taxonomy Explorer}
\label{subsec:taxonomy-explorer}
To demonstrate the practical use of the generated taxonomy, a prototype web
application, named Taxonomy Explorer, was developed. The application allows
users to navigate the five-level hierarchy, from Domains to Concepts, and
explore the organization of research areas in greater detail than the original
OpenAlex taxonomy.

\begin{figure}[!htbp]
    \centering
    \includegraphics[width=\textwidth,keepaspectratio]{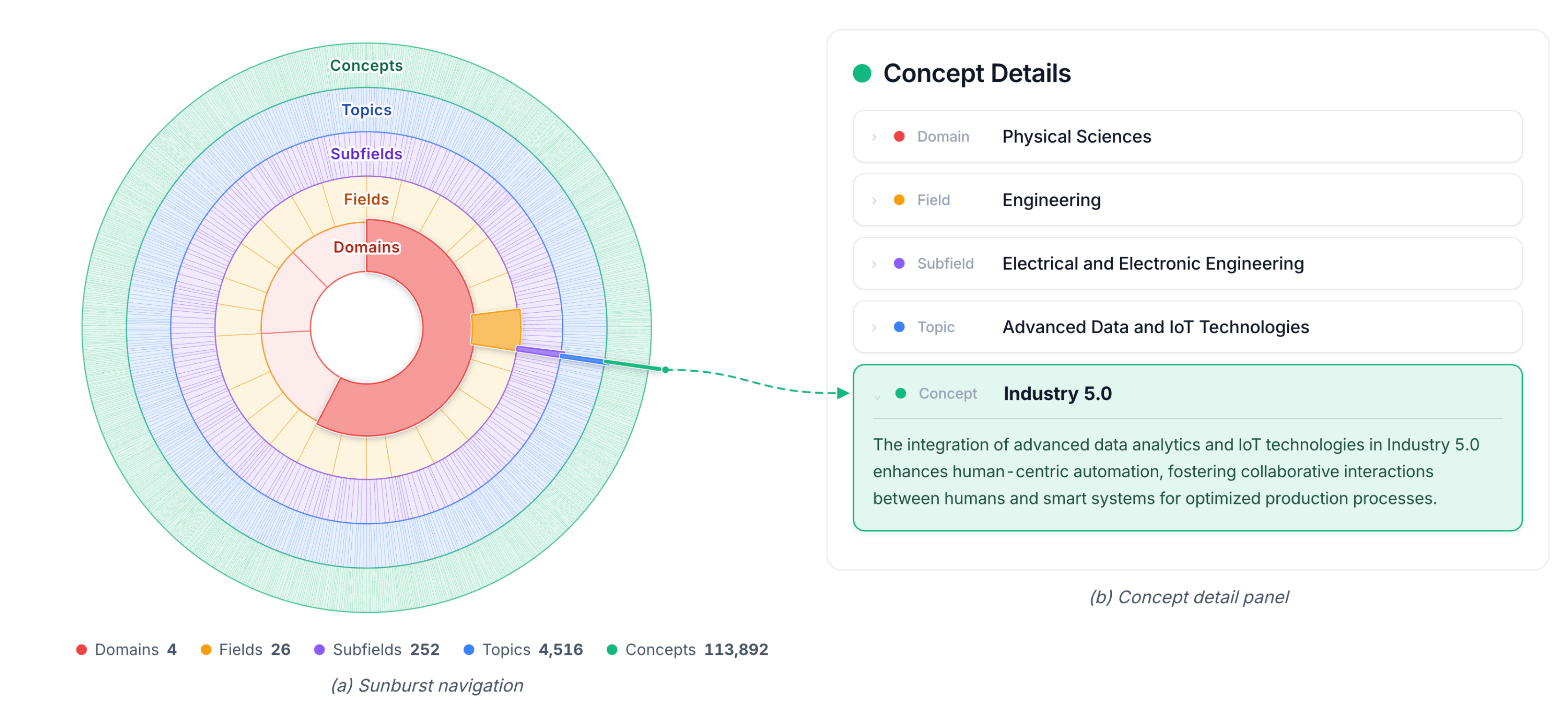}
    \caption{Interface of the Taxonomy Explorer prototype for navigating the five-level taxonomy from Domains to Concepts.}
    \label{fig:taxonomy-explorer}
\end{figure}

Fig.~\ref{fig:taxonomy-explorer} shows the Taxonomy Explorer interface. The
left panel provides a radial visualization of the hierarchy, allowing users to
move across Domains, Fields, Subfields, Topics, and Concepts. When users select
an item in the hierarchy, the right panel displays the corresponding taxonomic
path, together with the description and related Concepts.

In addition to hierarchical navigation, the Taxonomy Explorer provides search,
filtering, and export functions. 

A research prototype is available online, with access credentials available upon
request from the corresponding author.\footnote{\url{https://taxonomy.di.mdpi.ai/}}

\FloatBarrier

\subsubsection{Atlantis}
\label{subsec:atlantis}
Atlantis provides a semantic map of the Concept layer. In Atlantis, each point
corresponds to one Concept in a low-dimensional layout obtained from Concept
embeddings. The interface supports 2D and 3D views, search, and coloring by
Domain or Field. Selecting a Concept displays its label, explanation, and
hierarchical path, and retrieves papers tagged with that Concept.

The map is intended for navigation and inspection. Distances in the projected
space should not be interpreted as a formal measure of taxonomic relatedness.
Taxonomy Explorer and Atlantis therefore address different tasks: the former
exposes the explicit hierarchy, while the latter shows Concepts in embedding
space.

\begin{figure}[!htbp]
    \centering
    \includegraphics[width=\textwidth,keepaspectratio]{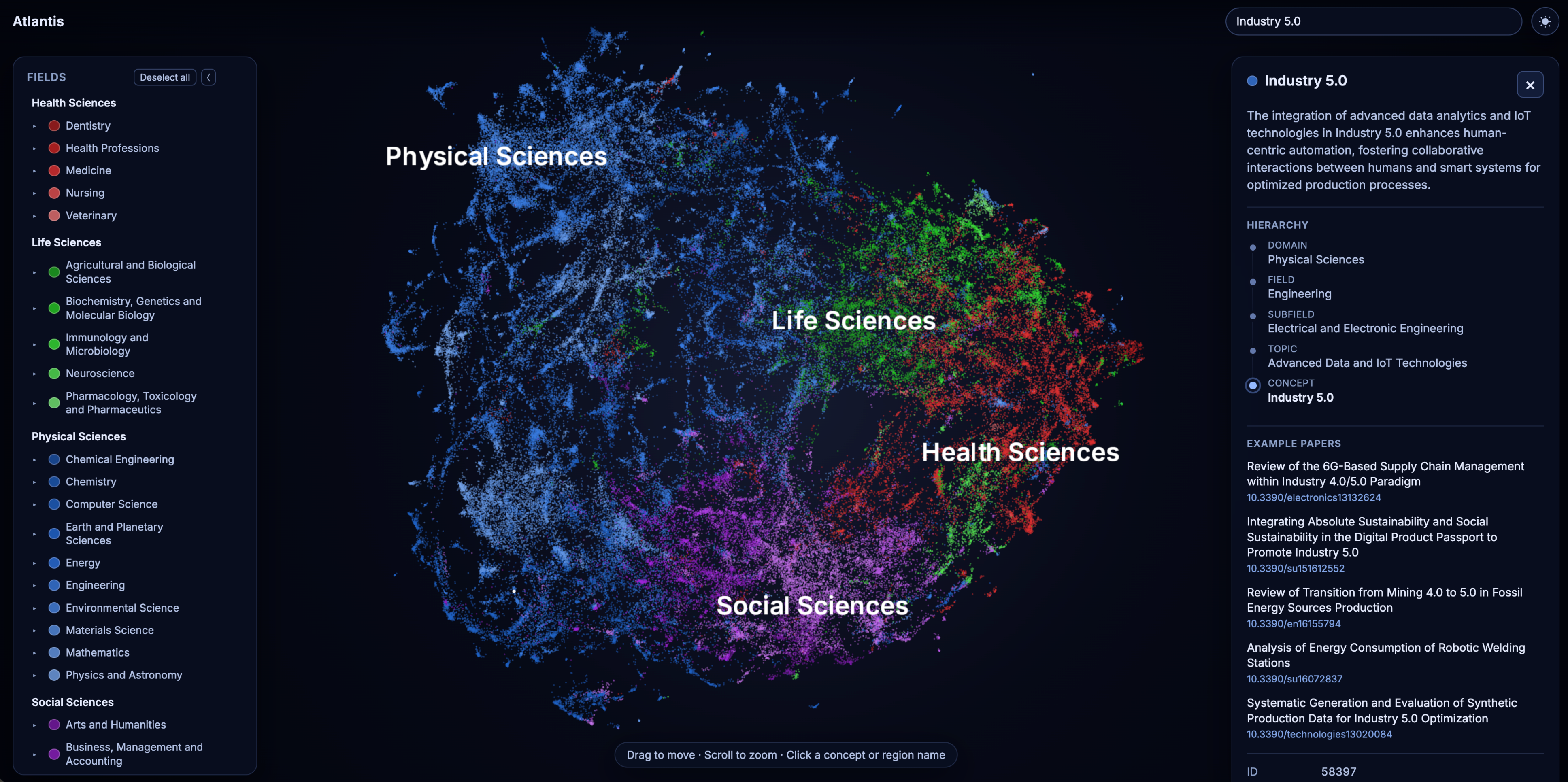}
    \caption{Interface of Atlantis. Each point is a Concept; selecting a Concept
    shows its taxonomic path and linked papers.}
    \label{fig:atlantis}
\end{figure}

Atlantis is publicly available without registration or
password.\footnote{\url{https://atlantis.mdpi.com}} Fig. \ref{fig:atlantis} shows the interface of Atlantis.

\FloatBarrier

\subsubsection{Open taxonomy dataset}
\label{subsec:open-dataset}
A versioned snapshot of the taxonomy is released as an open dataset on Hugging
Face.\footnote{\url{https://huggingface.co/datasets/mdpi-di/taxonomy}} The package
includes tables for Domains, Fields, Subfields, Topics, and Concepts, a flat
hierarchy view, Concept explanations, and stable identifiers. Levels 1--4 follow
the OpenAlex hierarchy; level 5 contains the Concepts produced by SCALE.
Explanations are short AI-generated descriptions intended for browsing and labeling,
not expert-curated definitions. The release contains taxonomy metadata only.
Paper full texts, abstracts, author lists, and human-evaluation microdata are
not included. The dataset is distributed under a CC0~1.0 license.

Table~\ref{tab:taxonomy-size} summarizes the released resource.
\begin{table}[!htbp]
\centering
\caption{Size of the released taxonomy snapshot.}
\label{tab:taxonomy-size}
\begin{tabular}{lr}
\toprule
Level & Count \\
\midrule
Domains & 4 \\
Fields & 26 \\
Subfields & 252 \\
Topics & 4,516 \\
\quad with $\geq 1$ Concept & 4,461 (98.8\%) \\
Concepts & 113,892 \\
\bottomrule
\end{tabular}
\end{table}

\subsection{Downstream Use and Expert Applicability}
\label{subsec:downstream-applicability}
Two additional forms of evidence are reported for the released taxonomy:
operational use in paper classification, and an expert evaluation of the
Concepts returned for sampled papers.

The taxonomy has been used in a production paper classification pipeline for the MDPI
corpus of nearly two million papers. 94\% of Concepts were
assigned to at least one paper. Papers received a median of 5 Concepts
(IQR 5--6).

A human evaluation was conducted with 22 domain experts across 11 fields
(110 papers; 220 paper$\times$evaluator units; 3,958 Concept judgments). For
each paper, experts inspected Concepts returned by two tagging configurations
based on different language models, GPT-5.4~mini and Qwen3 \citep{gpt54mini, yang2025qwen3}, and judged each
returned Concept as relevant or not relevant to the paper. Because production
tagging returns a median of five Concepts per paper, applicability is reported
as precision@5.

Across all units, mean precision@5 was 88.9\% for GPT-5.4~mini and 88.7\% for Qwen3. Precision@1
was 94.5\% and 95.9\%, respectively. Precision@5 varied by field
(Table~\ref{tab:applicability-by-field}), ranging from 75.0\% to 97.0\% for
GPT-5.4~mini and from 79.0\% to 98.0\% for Qwen3.

Agreement between the two experts assigned to the same field was examined on
Concepts rated by both. Observed agreement was $p_o = 0.76$. Under high
positive prevalence (TRUE rate $= 0.82$), prevalence-adjusted coefficients were
moderate (PABAK $= 0.52$, Gwet AC1 $= 0.66$), while Cohen's $\kappa$ was low
($0.19$) \citep{byrt1993bias, gwet2008computing, cohen1960coefficient}.

\begin{table}[!htbp]
\centering
\caption{Precision@5 by evaluation field (expert-study labels; not OpenAlex
Fields). Values are means over paper$\times$evaluator units. $n$ is the number
of papers per field. Overall is the mean over all paper$\times$evaluator units.}
\label{tab:applicability-by-field}
\begin{tabular}{lrrr}
\toprule
Field & $n$ & P@5(GPT) & P@5(Qwen) \\
\midrule
Physics & 10 & 75.0 & 85.0 \\
Materials Science 1 & 10 & 78.0 & 79.0 \\
Mathematics & 10 & 80.8 & 81.0 \\
Atmospheric/Environmental Science & 10 & 90.0 & 85.0 \\
Engineering & 10 & 90.0 & 88.0 \\
Botany/Plant science & 10 & 91.0 & 88.0 \\
Drug Development and Delivery & 10 & 91.0 & 95.0 \\
Molecular and Cellular Biology & 10 & 93.0 & 87.0 \\
Materials Science 2 & 10 & 95.0 & 93.0 \\
Biochemistry & 10 & 97.0 & 98.0 \\
History & 10 & 97.0 & 97.0 \\
\midrule
Overall & 110 & 88.9 & 88.7 \\
\bottomrule
\end{tabular}
\end{table}
\FloatBarrier

\section{Discussion}
\label{sec:discussion}
The results demonstrate the feasibility of constructing a fine-grained Concept
layer below OpenAlex Topics at scale. The selected configuration achieved 69.5\%
must-link precision and 87.5\% should-not-link precision, while the final
hierarchy associated at least one Concept with 98.8\% of OpenAlex Topics. These
findings support the broad hierarchical integration achieved by the framework,
although the lower must-link score indicates that some semantically related
candidate terms remain separated across communities.

\subsection{Scientific Contribution}
\label{subsec:scientific-contribution}
The main contribution of this work is the introduction of Concepts as a new unit
for organizing scholarly knowledge. Unlike individual author keywords, Concepts
consolidate semantically related candidate terms into named units located within
explicit disciplinary paths. They are designed to provide a finer-grained and
more reusable representation of specialized research areas than OpenAlex Topics.

Existing taxonomy construction approaches primarily organize terms or documents from text corpora, which makes them difficult to integrate with established scholarly knowledge organization systems \citep{zhang2018taxogen,lee2022taxocom,kargupta2025taxoadapt}. In contrast, SCALE extends OpenAlex through an automated taxonomy construction pipeline that discovers fine-grained Concepts and links them to the existing topic hierarchy.

SCALE also differs from the OpenAlex Keywords system \citep{openalexKeywords}. Whereas OpenAlex Keywords are derived from Topics and assigned to individual works, SCALE Concepts introduce a separate, finer-grained structural level that is constructed bottom-up from author keywords and integrated below the existing Topic hierarchy. This design enables SCALE Concepts to capture more specific research areas and to be updated as new scientific trends emerge.

SCALE helps reduce the gap between broad disciplinary classifications and highly specific document-level descriptions. The result is not only a quantitative extension of the taxonomy, but also the introduction of a conceptual layer designed to support more detailed analyses of the scientific literature. Concepts may therefore contribute to new applications in scientometrics and the monitoring of research evolution.


\subsection{From Taxonomic Tagging to Ontological Extension}
\label{subsec:taxonomic-tagging-ontological-extension}
A first operational application of SCALE is its use for systematically tagging articles published by MDPI and indexed in Scilit, an MDPI-developed scholarly discovery and indexing platform that aggregates metadata for scientific publications \citep{scilit}. In
operational use, papers are typically associated with around five Concepts,
consistent with the production default. The tagging pipeline is deployed across
the MDPI corpus of nearly two million papers. 94\% of Concepts appear in
at least one assignment, indicating that the taxonomy is already used broadly
rather than only in a narrow subset of branches. Large-scale tagging also
supports examination of how Concepts are distributed across disciplines,
journals, countries, institutions, and time periods, and thus monitoring of
research themes, emerging areas, and interdisciplinary connections in greater
detail than classifications based only on Topics.

Existing knowledge graph infrastructures may provide a natural environment for this extension. MarmotGraph, developed within EBRAINS, provides infrastructure for managing
metadata, controlled vocabularies, ontologies, and semantic relations across
scientific data \citep{ebrainsMarmotGraph}. From this perspective, Concepts tagged across Scilit publications could provide a broad scholarly layer to extend such graph-based representations beyond neuroscience and toward a more general organization of scientific knowledge.

At a later stage, this direction could support the evolution of the taxonomy toward a richer ontological representation. In the current implementation, Concepts are hierarchical taxonomic units rather than formally defined ontological entities. Future work could decompose Concepts into finer ontological entities and use citation patterns, co-occurrence signals, publication metadata, and other semantic evidence to infer explicit, interpretable relations among them.

Some limitations should also be addressed in the future. The must-link and should-not-link Concept pairs used for parameter selection were curated by the authors and may therefore introduce subjective bias, while the quality and conceptual coherence of sampled keyword clustering results would benefit from independent human evaluation. In addition, the current distribution of Concepts across Fields and Domains is uneven; although this may partly reflect differences in disciplinary breadth and publication activity, future work should investigate methods to mitigate this imbalance and improve coverage of underrepresented areas.


Overall, SCALE provides a structured and interpretable layer between OpenAlex
Topics and individual scientific documents. It supports a more fine-grained
organization of scholarly knowledge and offers a concrete basis for future
applications in article tagging, scientometric analysis, and ontology development.


\section*{Acknowledgements}
The authors thank MDPI for providing the AI infrastructure used in this work. The authors also thank Dietrich Rordorf and Luke Sui for their continuous financial support of this project. Finally, we thank the volunteer editors at MDPI who participated in the user evaluation and generously contributed their time and expertise.

\bibliographystyle{unsrtnat}
\bibliography{references}  






\end{document}